\documentclass[10pt,a4paper,conference]{IEEEtran}

\usepackage{alltt}

\usepackage{hyperref}

\usepackage{boxedminipage}
\usepackage{listings}

\usepackage{subfigure}

\newcommand{\code}[1]{{\relsize{-1}\textsf{#1}}}

\usepackage{graphicx} 

\usepackage{relsize}

\begin{document}

\title{Invertible Program Restructurings\\ for Continuing Modular Maintenance}

\author{
\IEEEauthorblockN{Julien Cohen}
\IEEEauthorblockA{ASCOLA\,team\,(EMN,\,INRIA,\,LINA)\\
University of Nantes\\ 
Nantes, France\\
Email:\,Julien.Cohen@univ-nantes.fr
}
\and
\IEEEauthorblockN{R\'emi Douence}
\IEEEauthorblockA{
ASCOLA\,team\,(EMN,\,INRIA,\,LINA)\\
INRIA\\
Nantes, France\\
Email:\,Remi.Douence@inria.fr}
\and
\IEEEauthorblockN{Akram Ajouli}
\IEEEauthorblockA{
ASCOLA\,team\,(EMN,\,INRIA,\,LINA)\\
EMN\\
Nantes, France\\
Email:\,Akram.Ajouli@mines-nantes.fr}
}
\maketitle

\begin{abstract}
When one chooses a main axis of structural decompostion for a software,
such as function- or data-oriented
decompositions, the other axes become secondary, which can be
harmful when one of these secondary axes becomes of main
importance.
This is called the tyranny of the dominant decomposition.

In the context of modular extension, this problem is known
as the Expression Problem and has found many solutions,
but few solutions have been proposed in a larger context of
modular maintenance.

We solve the tyranny of the dominant decomposition in
maintenance with invertible program transformations.
We illustrate this on the typical Expression Problem
example.
We also report our experiments with Java and Haskell
programs and discuss the open problems with our approach.
\end{abstract}



\section{Introduction}

Evolvability is a major criteria of quality for
enterprise software.
Evolvability is directly impacted by the
design choices on the
software architectures~\cite{Parnas1972}.
However, it is generally impossible to find software
architectures that are evolvable with respect to all
concerns.  So, one of these concerns has to be privileged at
the expense of other ones.
This is sometimes called the \emph{tyranny of the dominant
decomposition}~\cite{tyranny1999}.
At the micro-architecture level,
there are many ways to provide modular extensions which are
orthogonal to the main axis of decomposition of a code
structure, such as using open
classes~\cite{open-classes-2000} in which one can add methods 
without modifying the source code of those classes 
(see a review of several solutions in~\cite{Zenger-Odersky2005}).
However, these solutions generally break 
the regularity of the initial architecture (architectural degeneration), which results in a decrease in the maintainability 
(Sec.~\ref{sec-problem}).
This reveals a tension between modular
extension and modular maintenance.

In this paper, we use invertible program transformations
between pairs of ``dual'' code structures to solve the
tyranny of the dominant decomposition.
We illustrate this with two code structures, data- and
operation-oriented, for which we have built transformations with
refactoring tools (Sec.~\ref{sec-transformation}
and~\ref{sec-assessment}).
We also give the challenges to be solved to make this approach fully
automatic and scalable (Sec.~\ref{sec-challenges}),
based on our
experience with Java and Haskell
program transformations (Sec.~\ref{sec-implementation}).

\section{The Modular Maintenance Problem} 
\label{sec-problem}

In this section, we illustrate the fact that with fixed code
structures, maintenance cannot be modular with respect to
independent features (for instance, the set of operations on
a data type is independent of the set of possible cases in
that data type).
We illustrate this in an object oriented setting on a Java program, but the
problem is not restricted to object oriented architectures.

\subsection{Each Architecture Privileges Modular Maintenance on a Given Axis}
\label{sec-primary-decomposition}

When choosing a class structure (or more generally a module
structure) for a given program, one has to choose between
several possibilities with different advantages and
disadvantages~\cite{Parnas1972}.
We illustrate this with two possible class structures for a
simple evaluator which have dual advantages and
disadvantages : Composite (or Interpreter) and Visitor design patterns (Figs.~\ref{fig-java-composite}
and~\ref{fig-java-visitor}).
This program is the same that is often used to illustrate the
expression problem~\cite{expPb}, here given in Java.

\newcommand{\pfun}{P_{\mathit{fun}}}
\newcommand{\pdata}{P_{\mathit{data}}}

\lstset{frame=trbl, frameround=tfff, language=java, basicstyle=\sffamily \footnotesize}

\begin{figure}[htp]
\begin{lstlisting}
abstract class Expr {
  abstract Integer eval (); 
  abstract String show ();
}
\end{lstlisting}
\begin{lstlisting}
class Num extends Expr {
  int n;
  Num (int n){ this.n=n ;};
	
  Integer eval() { return n; }

  String  show() { return Integer.toString(n); }
}
\end{lstlisting}
\begin{lstlisting}
class Add extends Expr {
  Expr e1,e2 ;

  Add (Expr e1, Expr e2){
    this.e1 = e1 ;
    this.e2 = e2 ;
  }

  Integer eval() { return e1.eval() + e2.eval() ; }

  String  show() {
    return "(" + e1.show() + "+" + e2.show() + ")" ;}
}
\end{lstlisting}
\caption{Data decomposition (Composite/Interpreter pattern) in Java -- program $\pdata$.}
\label{fig-java-composite}
\end{figure}

\begin{figure}[htp]
\begin{lstlisting}
abstract class Expr {
  Integer eval(){return(accept(new EvalVisitor()));}

  String  show(){return(accept(new ShowVisitor()));}

  abstract <T> T accept (Visitor <T> v) ;
}
\end{lstlisting}
\begin{lstlisting}
class Num extends Expr {
  int n;
  Num (int n){ this.n=n ;};
	
  <T> T accept(Visitor <T> v){return v.visit(this);}
}
\end{lstlisting}
\begin{lstlisting}
class Add extends Expr {
  Expr e1,e2 ;

  Add (Expr e1, Expr e2){
    this.e1 = e1 ;
    this.e2 = e2 ;
  }
  <T> T accept(Visitor <T> v){return v.visit(this);}
} 
\end{lstlisting}
\begin{lstlisting}
abstract class Visitor <T> {
  abstract T visit(Num n);
  abstract T visit(Add a); 
}
\end{lstlisting}
\begin{lstlisting}
class EvalVisitor extends Visitor <Integer> {  
  Integer visit(Num a){ return a.n; }

  Integer visit(Add a) {
    return a.e1.accept(this) + a.e2.accept(this) ; }
 }
\end{lstlisting}
\begin{lstlisting}
class ShowVisitor extends Visitor <String> {
  String visit(Num a){return Integer.toString(a.n);}

  String visit(Add a) {
    return "(" + a.e1.accept(this) +
           "+" + a.e2.accept(this) + ")" ; }
}
\end{lstlisting}
\caption{Functional decomposition (Visitor pattern) in Java -- program $\pfun$.}
\label{fig-java-visitor}
\end{figure}

\begin{figure}[!htp]

\centering

\subfigure[Data decomposition (program $\pdata$).]{
\includegraphics[scale=1]{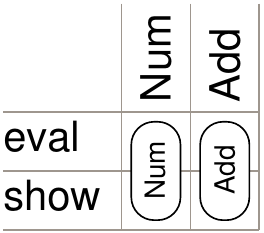} 
\label{fig-matrix-data}
}
\hspace{1cm}
\subfigure[Functional decomposition (program $\pfun$).]{
\includegraphics[scale=1]{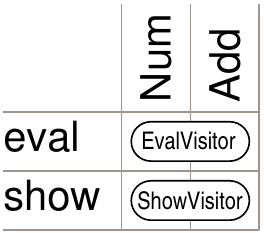}
\label{fig-matrix-fun}
}
\caption{Coverage of classes with respect to operations and data type.}
\label{fig-archis}
\end{figure}

The data type \code{Expr} represents the expression language
to be evaluated. It is represented by an abstract class. The
type \code{Expr} has a subtype for literals (\code{Num} for
integers) and another for an operator (\code{Add} for
additions).
Two operations (methods) are defined on the type \code{Expr}
: \code{eval} to evaluate expressions and \code{show} to
transform them into strings. Their behavior is defined by
case on subtypes.
We call the code that defines the behavior of these
two operations the \emph{business code}.
In the following, we are interested in the location of the
business code in the class structure (which determines the
modularity of maintenance tasks).

In the Composite architecture
(Fig.~\ref{fig-java-composite}), the business code which deals
with a given subtype is delimited by the corresponding
class.
The diagram in~Fig.~\ref{fig-matrix-data}
shows a matrix indexed on subtypes and operations. 
The concrete classes form a partition of the matrix according to the
subtypes covered by the business code they contain.
For instance, the class \code{Add} contains the business
code for the two operations but only the part which concerns the
subtype \code{Add}.

In this architecture, the maintenance concerning a given
subtype is modular: when the requirements or the internal
representation of a subtype changes, all the changes in the
business code are located in the corresponding class.
On the other hand, the maintenance of a given operation is not
modular: when the requirements for an operation changes, the
changes in the business code can be spread over the
subclasses.

The program with the Visitor architecture
(Fig.~\ref{fig-java-visitor}) has \emph{dual properties} with
respect to modularity.
Its class structure makes that all the business code
related to a given operation are located in a single class.
For instance, the class \code{EvalVisitor} contains all
the business code for the method~\code{eval}.
The matrix of Fig.~\ref{fig-matrix-fun} pictures that the
classes with the business code do not cover subtypes anymore
but operations.

In this architecture, the maintenance of a given operation is
modular: when the requirements for an operation changes, all
the changes in the business code are located in a single class.
On the other hand, the maintenance of a given subtype is not
modular: when a subtype changes, the
changes in the business code can be spread over the
visitor classes.

This duality illustrates the tyranny of the dominant
decomposition in action: whatever program structure is
chosen, some maintenance will be non modular.
In the following, we call this the \emph{modular maintenance
  problem}.

\subsection{The Modular Maintenance Problem: Functional Programming Style}

The opposition between data oriented architectures and
operation oriented architectures is not specific to object
oriented programs.
In functional languages, functions are frequently defined by
pattern matching on the structure of data.
This corresponds to an operation oriented architecture: 
maintaining existing functions is modular
but maintaining an existing case in the data type
is not modular (the changes in business code can be
spread over several functions).

An alternative way to define functions is to use traversal
operators (\emph{fold} catamorphisms) which take as
parameter one function for each case in the data type. Since
these parameter functions are specialized for given
cases, it is relevant to group them into modules containing
business code for specific cases of the data type.
This corresponds to a data oriented architecture: 
maintaining a case in the data type is
modular but
maintaining a function 
is not modular (the changes in
the business code are spread over several modules)~\cite{CohenDouence2011}.

\subsection{Modular Extensibility (The Expression Problem)}
\label{sec-expression-problem}

A problem closely related to the modular maintenance problem
exists with extensions:
in the Composite architecture (we return to an object
oriented setting), \emph{adding} a new subtype is modular (the
business code is added in the new class) but adding a new operation
is not (the business code is spread over several classes),
and inversely in the Visitor architecture.
This is known as the \emph{Expression Problem}~\cite{expPb}.

\begin{figure}[!htp]

\centering

\subfigure[Data oriented initial decomposition.]{
\hspace{3mm}
\includegraphics[scale=0.9]{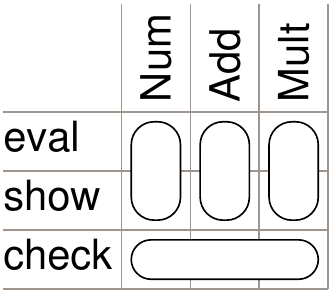}
\hspace{3mm}
\label{expr-prob-erosion-data}
}
\hfill
\subfigure[Operation oriented initial decomposition.]{
\hspace{3mm}
\includegraphics[scale=0.9]{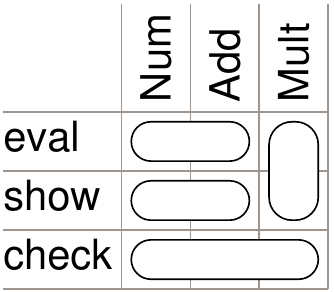}
\hspace{3mm}
\label{expr-prob-erosion-fun}
}
\caption{Architecture after two modular extensions.
We consider the two initial architectures described before
(Fig.~\ref{fig-archis}), extended with a subtype named
\code{Mult}, then extended with an operation named \code{check}.
}
\label{expr-prob-erosion}
\end{figure}

There are many ways to extend the data-type or the
set of operations indifferently in a modular way
(see~\cite{Zenger-Odersky2005} for a review of some
solutions).
However, after the modular addition of an operation, the code is not modular anymore with respect to subtypes (see Fig~\ref{expr-prob-erosion}), and after the modular addition of a subtype, the code is not modular anymore with respect to operations.
For this reason, (language-based) solutions for modular
extension conflict with modular maintenance.

\section{Invertible Program Transformations to solve the Modular Maintenance Problem}
\label{sec-transformation}
\label{sec-solution}

Chains of refactoring operations can be used to change the
structure of programs while preserving their external
behavior~\cite{Griswold1991}.
We propose to use \emph{invertible} chains of refactoring
operations to solve the problems of modular maintenance and
modular extension.

\newcommand{\Tdf}{\ensuremath{T_{DF}}}
\newcommand{\Tfd}{\ensuremath{T_{FD}}}

First, the two programs $\pdata$ and $\pfun$
of the previous section can be transformed one into the other by a
behavior preserving program transformation, and inversely
(we have implemented such invertible transformations for
Java and for its Haskell functional counterpart, see
Sec.~\ref{sec-implementation}).

\begin{figure}[htp]
\centering
~\hspace{3mm}\includegraphics[scale=0.7]{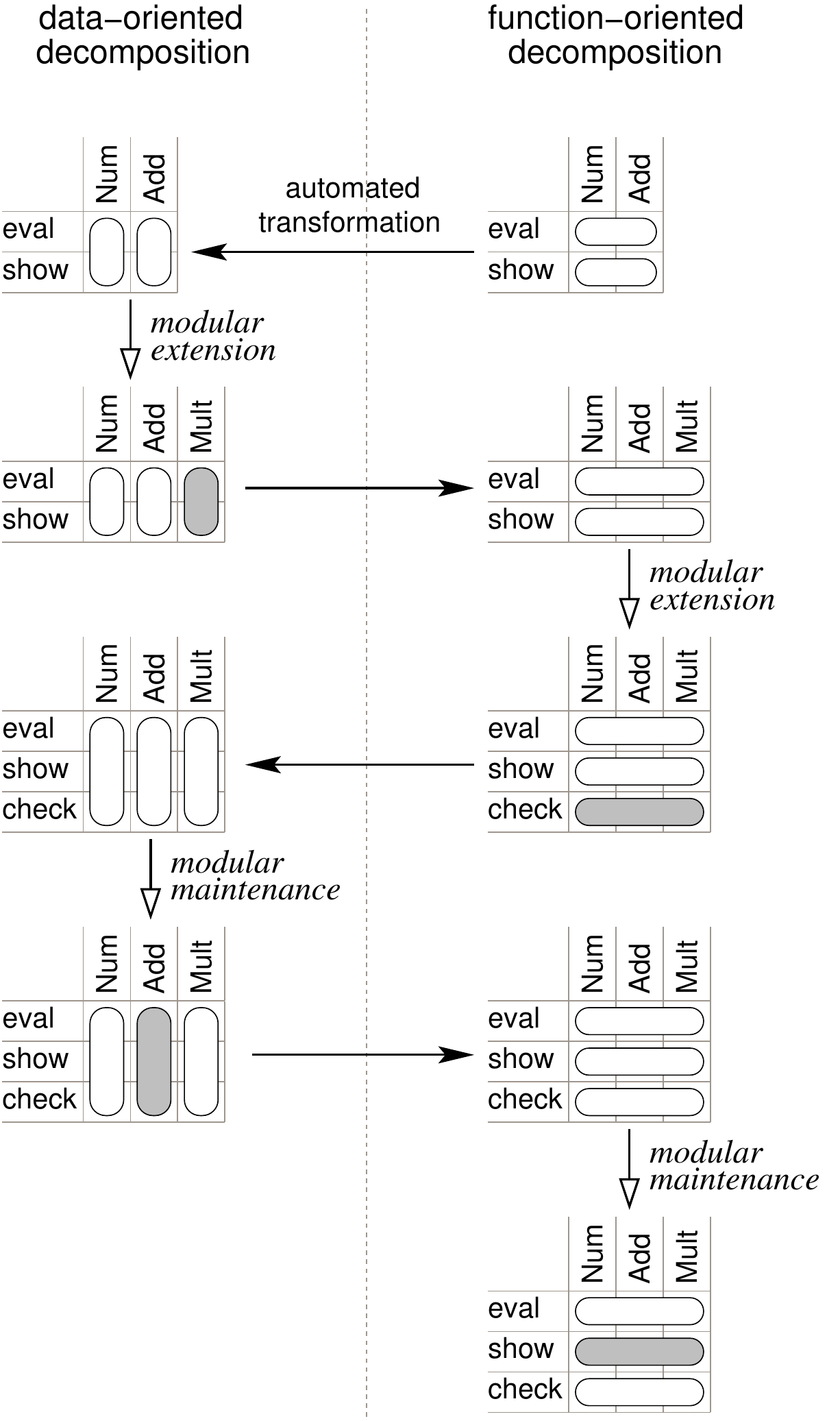}
\caption{Scenario for 4 evolutions with architecture transformations. The initial code is extended with the subtype \code{Mult} and with the operation
 \code{check}, then maintenance tasks are performed on the subtype \code{Add} and on the operation \code{show}. Structure transformations are performed so that all the evolutions are modular.}
\label{fig-ideal}
\end{figure}

Such transformations solve the problem of modular
maintenance: when one faces an evolution task (which
requires either to add a new subtype/operation or to modify an
existing subtype/operation) to be performed which is not
modular in the available form of the program, he applies the
convenient transformation to get the program into the
convenient form, then he implements the evolution in a
modular way (see Fig~\ref{fig-ideal}).
In the case of an extension, the resulting architecture is
not degenerated.

Once the evolution is implemented, one can either leave the
program in the last form, or apply the
inverse transformation to recover the initial structure with
the implemented changes propagated.

\section{Experiment: Implementation of Architecture Transformations with Refactoring tools}
\label{sec-implementation}
We have made conclusive experiments with Java and Haskell.

In Java, we have put to test Composite $\leftrightarrow$ Visitor
transformations with Eclipse~\cite{eclipse-web} and IntelliJ \textsc{idea}~\cite{intellij-web} refactoring
tools.
We describe in~\cite{CohenAjouli2011} the abstract
algorithms we use, some variants we propose and the
specificities due to the use of these tools.
The whole transformation is not automated yet (we plan to
automate this algorithm by using the tools API in
conjunction with pattern detection tools such
as~\cite{Smith03spqr}).

In Haskell, we have performed transformations between
function oriented and data oriented architectures with the
Haskell Refactorer~\cite{HaRe-Wrangler2008}.
We describe in~\cite{CohenDouence2011} the abstract
algorithms we have designed.
The transformations are automated for several examples of
programs. They are concretely defined by scripts much of
which is reusable for other programs.
We have customized the API of the Haskell Refactorer to be
able to automate the transformation steps
(see~\cite{CohenDouence2011, view-switcher-web}).

\subsection{Results}

Here is what we have observed from our experiments:

\begin{itemize}

\item
 The external behavior is preserved by transformations, as well as
 type safety. 

\item
 We find back the initial source code after performing a
 transformation and its inverse, except for the layout and
 the comments which have been disturbed.

\item In the Java experiment, the visibility for the composite class elements
 has to change when
 passing from Composite to Visitor
 structures.
 This is not related to the transformation but rather 
 to the nature of the Visitor pattern.

\item
 On small/medium-size programs (we used programs with 6 subtypes and 6
 operations), Java refactoring tools were fast enough, while
 the Haskell Refactorer was very slow:
 the Composite$\rightarrow$Visitor takes about 3
 minutes (plus several hours to chain the operations manually) 
 while the Haskell Refactorer could take 30 seconds for an elementary
 renaming (but transformations are automated).

\item 
Our algorithms are sensitive to variations in the initial
structure.

\item
A few refactoring operations were needed but not covered by
the tools.
For Java, we have made some refactoring
steps manually to validate the transformation algorithms.
For Haskell, we have added five operations into the
tool to be able to automate the full transformations.

\end{itemize}

\section{Assessment}
\label{sec-assessment}
\label{assessment}

The results above show that our proposal is workable only
with efficient tools.
We expose the challenges to be solved to provide such tools
in Section~\ref{sec-challenges}. 
In the rest of this section, we discuss more generally the
pros and cons of our proposal.

Our approach
does not rely on a particular programming language
(we have dealt with two different languages).
It applies as soon as two alternative programming structures can be
expressed in a language.
It results that:\begin{itemize}

\item Our solution can be applied to legacy systems.

\item 
The programmer's skill in the programming language is
 sufficient to implement modular evolutions.
Our approach does not require that the programmer should
master specific composition mechanisms such as aspects,
mixins, open classes, or hyper-slices.

\item Our solution does not induce runtime overhead.

\end{itemize}

On the other hand, a transformation tool capable of
performing the architecture transformation must be available
for the considered language (see Sec.~\ref{sec-implementation-with-refactorings}).

Our solution is not limited to the data-centered 
\textit{versus} the function-centered structures (see
 Sec.~\ref{sec-paired-transformations}). 
It is not even limited to two structures (with the limitation
that for each new structure to be considered, a
pair of transformations must be available or defined).

Last, programmers already familiar with the initial program
structure may lose their marks in a second structure.

\section{Challenges for Tool Support}
\label{sec-challenges}

\label{sec-implementation-with-refactorings}

Using refactoring tools to implement architectures
transformations make transformations
 easy to design and tune since refactoring operations are rather 
high-level transformations. 
Refactoring operations are also easily composed to make more
complex operations that can be used as components for
building our transformations. Moreover, chains of
refactoring operations are already used to describe the
introduction of design patterns into existing
code~\cite{Kerievsky:2004}.

On the other hand, other aspects of refactoring tools make
their use not entirely satisfactory in our context. 
We now
discuss the challenges to get over in order to make our
solution of industrial strength.

\subsection{Soundness.} 

Using refactoring tools to implement architectures
transformations has the advantage that the soundness of the
transformation relies on the refactoring tool.
However, it is frequent to face bugs in refactoring tools
(we have faced several bugs in refactoring tools during our
experiments).
A single bug in a chain of elementary transformations 
make all the process fail.

Proofs of correctness of refactoring
operations exist~\cite{formalJavaRefactoring2006,
  Sultana-Thompson2008},
but we cannot expect refactoring tools to be proven correct
in near future.
However, we can expect that popular refactoring tools
progressively become safer when bugs are reported.

\subsection{Layout Preservation, Invertibility}
\label{sec-layout}

With current refactoring tools, it seems
impossible to design invertible architecture transformations
that take layout and comments into account.
A solution is to provide invertible versions of refactoring
operations within the meaning of Bohannon \emph{et al.}~\cite{Bohannon:2008}:
%
non invertible operations, such as deletion, can become
invertible by keeping a trace of the program
before transformation.
This suggests that it could be useful to keep a reference
architecture and to use alternate ones only
temporarily.
That would also allow several maintainers to share a common
reference model in the case of teamwork.

\subsection{Speed and Flexibility}
\label{sec-efficiency}
\label{sec-variability}

To be workable, our proposal must be automated with convenient tools.
The underlying refactoring tool must be
sufficiently fast (which is the case for popular tools such
as Eclipse but not for academic, prototype tools such as the
Haskell Refactorer).

Moreover, to avoid time-consuming user interactions, transformation
tools must be capable of detecting  structures in
programs and of adapting the chain of refactoring operations to
these structures.
We can consider using pattern detection tools bearing
variations in pattern instances (such as~\cite{Smith03spqr}) and
either to adapt the chain of refactoring operations to these
variations upstream or to use tools that infer such chains of
refactoring operations.
For instance, in~\cite{reconstruction-of-refactorings-2010},
the target structure is described by logic constraints.

\subsection{Failures and Pre-Conditions}

Since each operation of the chain of refactorings requires some
preconditions to be satisfied, it may occur that the user is
advised that the transformation cannot be achieved only
during the transformation process.
For this reason, providing pre-conditions for our
transformations is desirable (pre-conditions for chains of
refactoring operations are explored in Kniesel and
Koch~\cite{composition-of-refactorings2004}).

\subsection{Macro-Architectures} 

In this paper, we have dealt with source-code level
architectures (micro-architectures).
But alternate structures are also useful at the system
level (macro-architectures)~\cite{views-components-2010}.
This suggests that transformations between dual
macro-architectures as well as refactoring tools for
composition/coordination languages should be explored.

\section{Related Work}
\label{related-work}

\subsection{Program Restructuring and Refactoring to Patterns}

Work on refactoring have always considered that the aim of
refactoring is to improve code structure (and so 
evolvability)~\cite{opdyke92, Griswold1993}.
Since most of that work takes place in an object-oriented
context, it is natural that design patterns have been
considered as target code structures~\cite{Cinneide:2000, Kerievsky:2004}. 
Switching to alternate patterns has also been considered
recently~\cite{Hills:2011}.

All that work is a basis for our proposal, but we are more
demanding: we need invertible transformations, full automation, etc. (see Sec.~\ref{sec-challenges}).

\subsection{Views}

Offering alternate views of software artifacts is not a new
idea and is useful in
practice~\cite{empirical-study-multiple-view-1992}.

Wadler proposes a concept of views that allows to handle
datatypes with several interfaces for pattern
matching~\cite{Wadler:1987}.
This permits the programmer to use the more convenient
interface to implement an algorithm so that its design and
evolvability are improved.
However, extension of the data-type still requires
cross-cutting changes in the algorithms.
Also, the underlying mechanisms can introduce a run-time overhead.

Tarr \emph{et al.}~\cite{tyranny1999} propose to construct
programs by composing possibly overlapping compilation units
(\emph{hyperslices}), each describing a concern.
Hyperslices are useful for program comprehension since the
concerns are clearly separated, but since they can be
overlapping, evolutions can be difficult to implement.

Mens \emph{et al.}~\cite{Mens:2002} propose a system where
concerns are described by a set of properties
(a \emph{view}). As for Tarr \emph{et
al.}~\cite{tyranny1999}, these views help for program
comprehension and help to check that an evolution does not
violate the properties of a concern, but it does not make
the evolution modular.

Shonle \emph{et al.}~\cite{Shonle2007} also allow to define
patterns (views) describing crosscutting parts of code of
interest, but in addition the programmer can implement
concern-specific evolutions based on these patterns.

We share with Black and Jones~\cite{Black2004} a same
theoretical concept of views: alternate forms of a program
which are computed from that program, which external
behavior are equivalent, with different structural
properties and that can be transformed back to the initial
structure.
However, whereas we defend the use of ``dual'' code
structures expressible in a same language, they propose to
use language extensions to support alternate code structures.
For instance, whereas we propose the Visitor code structure
as a function oriented alternate view for the data-oriented
 code structure, they prefer to use a flattened
class hierarchy (expressed in an extension of
the initial language) so that all the business code
for a given operation is grouped.

The number of proposals for concepts of views shows that
there is an inclination to provide multiple views of
software artifacts to improve separation of concerns.
However, the work cited in this section have a common
property: they are built \emph{on top} on existing languages
(language extensions, pattern languages, additional
composition mechanisms...).
This means that the programmer must be skilled not only in
the base programming language, but also in the technology
that provides views (to understand, use, define, modify or
compose views).
We stand out from this by not requiring these skills but
by requiring that convenient transformations are provided instead.

\subsection{Transformations between other pairs of dual architectures}
\label{sec-paired-transformations}

Our approach is not limited to function oriented versus data oriented
views.
First,  one can also
provide a security view, a transaction view, or any view
which reifies a concern that is subject to change.
Second, views can be used to other aims than modularity. It
can be used to navigate between conflicting design choices.

\subsubsection{Add or remove structure}

For instance, instead of changing the main axis of
structure, one can  need to add/remove
structure.
Adding a function that factorizes some code allows to hide a
behavior, to name a concept, to remove code duplicates,
to move piece of code for
a concern to a given module/class.
On the opposite, inlining/unfolding a function enables to remove an indirection or
a dependency to a module, to
ease an analysis.
The same is true for class hierarchies (class hierarchies
make clean architectures but behavior code is spread over several files),
or for aspects (understanding aspect interactions can be
tricky).
Is is also sometimes useful to add/remove polymorphism or
machinery such as iterators to improve understanding and analysis.

\subsubsection{Change internal behavior}
More generally, software engineering offers fundamental
design choices that could be (at least partially) supported
by views.
For instance,
$\lambda$-lifting~\cite{Johnsson85lambdalifting}
(resp. \mbox{$\lambda$-dropping} \cite{Danvy:1997:LTR:258993.259007})
adds (resp. removes) extra function parameters corresponding
to free variables.  The $\lambda$-lifted view 
 promotes function reuse, and the
$\lambda$-dropped view 
 promotes efficiency. 
A same relationship exists between continuation passing
style and direct style~\cite{190866}.

Another design tradeoff exists between computation time and storage
in memory.
This is exemplified by the choice to use memoization.
Second example: when implementing a collection, one has the choice to
compute the number of elements in the data-structure on
demand or to store it in the data-structure and maintain it.
In the latter case, yet another tradeoff occurs between updating the
stored size at each update of the elements, or updating it
only when the size is accessed.

Finally, a last tradeoff is related to when a computation
occurs. For example, two processes can communicate 
synchronously or asynchronously, with or without buffers, etc.

These views are quite general and maybe impossible to support
automatically.
But, when possible, views
can reduce the impact of making these design choices early, when future changes in requirements are not
known yet.

\section{Conclusion}

The contributions of this article are the following:
\begin{itemize}

\item We show how invertible program transformations make
  continual modular maintenance along crosscutting concerns
  feasible.

\item We point some technical and scientific challenges to
  make the approach workable, based on our experience in
  building tools to support such transformations.

\end{itemize}

Applying invertible structure transformations
 with (yet to provide) appropriate, fully automatic tools
can enable to:

\begin{itemize}

\item Reduce structure degeneration with continual
  change.

\item Reduce the impact of early design choices and 
  reduce the cost of maintenance or incremental development for
  concerns which are transverse to the main axis of
  decomposition.

\item Reduce the need for specific programming skills (such
  as aspects) for separation of concerns.

\end{itemize}

\section*{Acknowledgment}
The authors would like to thank Jean-Louis Giavitto (\mbox{IRCAM}),
Jean-Claude Royer (EMN) and the Haskell \mbox{Refactorer}  team
for their comments.

\bibliographystyle{IEEEtran}
\bibliography{biblio}  

\end{document}